\documentclass[ reprint,2,amsmath,amssymb,aps]{revtex4-1}

\usepackage{graphicx}
\usepackage{dcolumn}
\usepackage{bm}
\usepackage{natbib}
\usepackage[table]{xcolor}
\usepackage{tabularx}
\usepackage{array}
\usepackage{colortbl}

\begin{document}

\preprint{APS/123-QE}

\title{Quantum storage and manipulation of heralded single photons in atomic quantum memories}
\author{Pin-Ju Tsai$^{1,2}$, Ya-Fen Hsiao$^{2,3}$, and Ying-Cheng Chen$^{2}$}
\email{chenyc@pub.iams.sinica.edu.tw}
\affiliation{$^{1}$Department of Physics, National Taiwan University, Taipei 10617, Taiwan
\\
$^{2}$Institute of Atomic and Molecular Sciences, Academia Sinica, Taipei 10617, Taiwan
\\
$^{3}$Molecular Science Technology Program, Taiwan International Graduate Program, Academia Sinica and National Central University, Taiwan}
\date{\today}
\begin{abstract}
We demonstrate the storage and manipulation of narrowband heralded single photons from a cavity-enhanced spontaneous parametric downconversion (SPDC) source in the atomic quantum memory based on electromagnetically induced transparency. We show that nonclassical correlations are preserved between the heralding and the retrieved photons after storage process. By varying the intensity of the coupling field during retrieval process, we further demonstrate that the waveform or bandwidth of the single photons can be manipulated and the nonclassical correlation between the photon pairs can be even enhanced. Unlike previous works, our SPDC source is single mode in frequency, which not only reduces the experimental complexity arising from external filtering but also increases the useful photon generation rate. Our results can be scaled up with ease and thus lay the foundation for future realization of large-scale applications in quantum information processing.
\end{abstract}
\pacs{32.80.Qk, 42.50.Gy}
\maketitle
Quantum memories are devices that can store and retrieve photonic quantum states on demand\cite{QM_review1}. The ability of quantum memories to synchronize probabilistic events makes them a key component in quantum repeaters for long-distance quantum communication\cite{QR_review}, linear-optics-based quantum computation\cite{LOQC_review} and enhancing the multiphoton generation rate\cite{Enhance_MP}. Over the past two decades, intensive efforts and significant progress have been made in the development of high-performance quantum memory\cite{QM_review2}. While many of these works used the weak coherent laser pulses, some recent experiments have demonstrated the quantum storage of true single photons\cite{Akiba2009,Clausen2011,Zhang2011,Zhou2012, Rielander2014,England2015,Distante2017,Wang2019}. For the realization of quantum memories and their applications, single photons with suitable properties for storage are crucial. Single photons can be produced on demand by single atoms inside a cavity\cite{CavitySinglePhoton1,CavitySinglePhoton2,CavitySinglePhoton3} or in a heralded way through spontaneous Raman transitions\cite{RamanSinglePhoton} or spontaneous four-wave mixing\cite{FWMSinglePhoton1,FWMSinglePhoton2,FWMSinglePhoton3,FWMSinglePhoton4} in atomic ensembles. One advantage of the single photons generated from atoms is that their frequencies are right on the atomic transition and their bandwidths can easily be narrower than the linewidth of the atomic transition, allowing efficient interaction between photons and atoms. However, the disadvantage of such single-photon sources is the relatively complicated setup, which is cumbersome to scale up for large-scale experiments.   

Photon pairs generated from SPDC process with nonlinear crystals are widely used in quantum information processing. Heralded single photons can be produced from such photon pairs in which the detection of one (idler) photon heralds the production of its twin (signal photon). Applications of SPDC-based heralded single-photon storage to quantum repeaters\cite{PhotonPairRepeater} or to enhance the multiphoton rate\cite{Enhance_MP} have been proposed. To reduce the broad bandwidth (typically hundreds of GHz) for single photons generated by SPDC, cavity-enhanced SPDC has been developed to generate photon pairs with a bandwidth of the MHz level\cite{CavitySPDC1,CavitySPDC2,CavitySPDC3,CavitySPDC4,CavitySPDC5,CavitySPDC6,QST_photon_source}. The storage of such heralded single photons in quantum memories based on atomic ensembles\cite{Akiba2009,Zhang2011} and rare-earth-doped crystals have been demonstrated\cite{Clausen2011,Rielander2014}. However, the photon sources used in all of these works were multimode in frequency. External etalon filters stabilized to the atomic transition frequency are needed to filter out the unwanted frequency modes. This not only complicates the experimental setup but also reduces the useful photon generation rate. In a recent work\cite{QST_photon_source}, we developed a narrowband, single-mode photon-pair source by cavity-enhanced SPDC. Based on this, we demonstrate the quantum storage and manipulation of heralded single photons in atomic quantum memories based on electromagnetically induced transparency (EIT) with cold atoms. Nonclassical correlation between the heralding and retrieved photons is demonstrated. By varying the coupling field intensity during retrieval, the single-photon waveform or bandwidth can be manipulated and the nonclassical feature can be even enhanced. We have obtained a retrieval efficiency of up to 36\%, which is currently limited by the bandwidth of the single photons, the optical depth of the medium, and the coupling-field-induced decoherence due to the off-resonant excitation\cite{YF2018}. Future improvements of the storage efficiency are possible\cite{YF2018}. Our demonstration of quantum storage and manipulation of heralded single photons from a compact cavity-SPDC source can be scaled up with ease and thus lays the foundation for future realization of large-scale protocols in quantum information science.     

The experimental setup is shown in Fig.\ref{setup}(a). It is composed of a photon source and a cold atom system. In the photon source, nondegenerate narrowband photon pairs are generated through cavity-enhanced SPDC. We briefly mention essential points here and more details are shown in\cite{sup, QST_photon_source}. The pump beam is from an external cavity diode laser with a wavelength of 407 nm. After passing through a focusing lens $L_1$, it pumps the type-II periodically poled KTiOPO4 (PPKTP) crystal. The idler photons, with a wavelength of $\sim$ 780 nm, acts as trigger to herald the generation of signal photons. The wavelength of the signal photons is locked to the cesium $D_2$-line transition of $|6S_{1/2},F=3\rangle\rightarrow|6P_{3/2},F=4\rangle$. The bandwidth of the photon pairs is $\sim$ 6 MHz, comparable to the natural-linewidth of cesium $D_2$-line of $5.23$MHz. The clustering effect of type II phase matching and the double-pass pumping results in single-longitudinal-mode operation of the photon pairs\cite{Chuu_theory,CavitySPDC3,QST_photon_source}. Therefore, there is no need for an external filter to filter out the unwanted frequency mode. The idler photons are coupled into a 0.5m single-mode fiber and then detected by single photon counting modules (SPCM$_i$). The signal photons pass through a 150m long polarization-maintaining fiber to reach the quantum memory laboratory. 

\begin{figure}
\includegraphics[width=8 cm,viewport=165 80 640 530,clip]{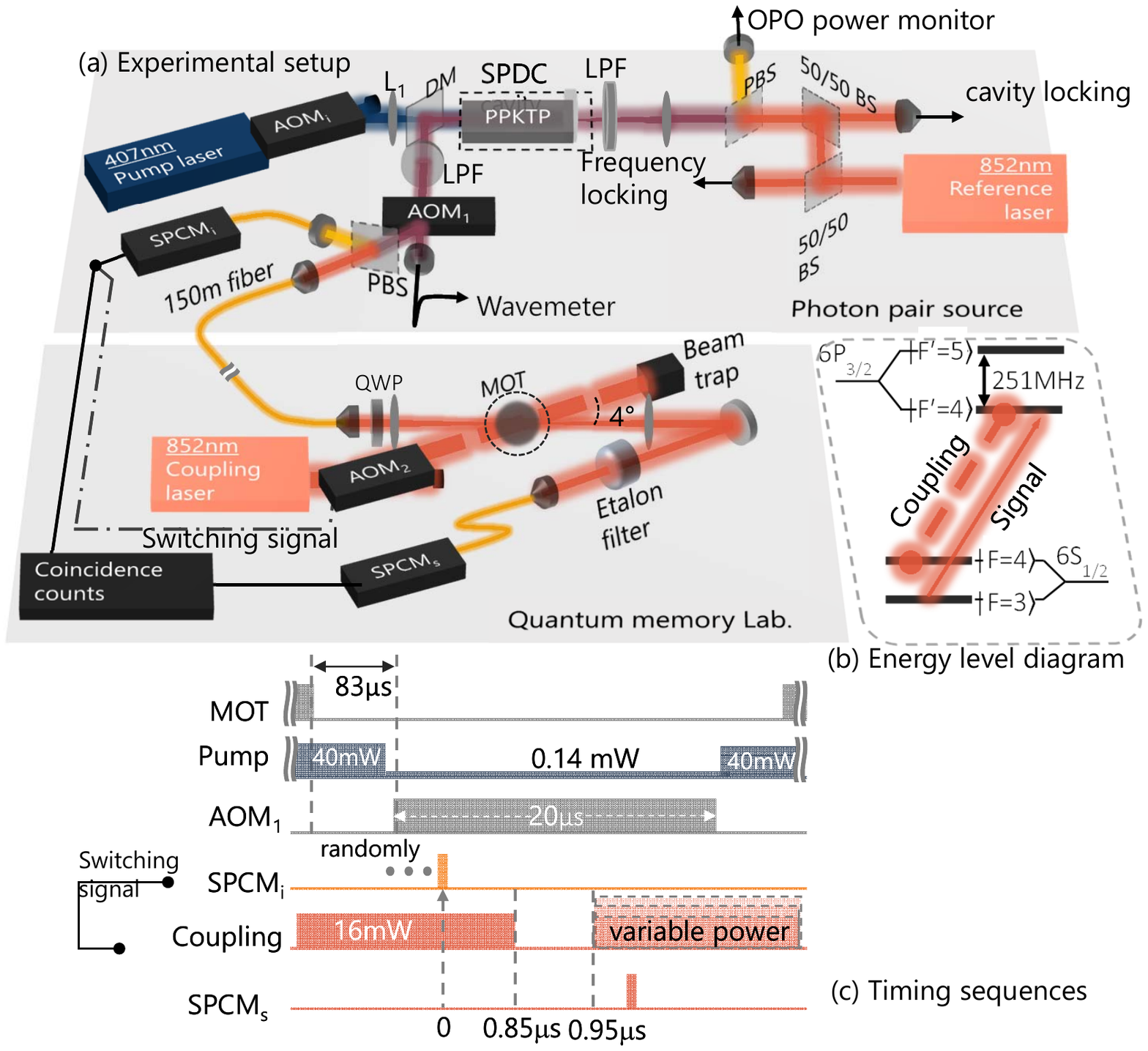}
\caption{(a) The experimental setup comprises two parts: the cavity-enhanced SPDC photon pair source and the cold atomic quantum memory based on EIT.  L, lens; QWP, quarter wave plate; BS, beam splitter; PBS, polarizing beam splitter; LPF, long pass filter; DM, dichroic mirror. (b) Relevant energy levels of $^{133}$Cs atom and laser excitations. (c) Timing sequences of the experiment. The storage process is completed by switching the coupling beam off and on upon triggering from the idler photons.} 
\label{setup}
\end{figure}

Our atomic quantum memory is based on a magneto-optical trap (MOT) of cesium\cite{YFMOT2018}. We typically trap $\sim 4\times10^8$ cold atoms with a temperature of 180 $\mu K$. To increase the optical depth of the atomic ensemble, we employ a temporally dark MOT and the Zeeman-state optical pumping\cite{highOD}. The optical depth of the atomic clouds is around 55$\pm$10. A coupling beam drives the transition of $|6S_{1/2},F=4\rangle\rightarrow|6P_{3/2},F=4\rangle$ and intersects with the signal photons at an angle of $\sim4^{\circ}$. As shown in Fig.\ref{setup}(b), the coupling and signal transitions form a $\Lambda$-type EIT system, which serves as a memory platform for the signal photons. By switching off the coupling field, the quantum state of the signal photons can be written into the collective ground-state coherence of the atomic ensembles\cite{EITmemory2000}. Moreover, this EIT-memory provides an avenue for manipulation of the stored signal photons by configuring the retrieval process\cite{HakutaPRA,Yu2005,Yu2006}. 

Before employing the heralded single photons for storage, we first characterize the properties of the photon pairs by measuring the Glauber correlation function (or biphoton wavefunction) $G_{s,i}^{(2)}(\tau)$, which is defined by $\left \langle \hat{a}_i^{\dagger}(t+\tau)\hat{a}_s^{\dagger}(t)\hat{a}_s^{\dagger}(t)\hat{a}_i(t+\tau) \right \rangle$, where $\tau$ is the time delay between the arrival of the signal and idler photons at the detector. The biphoton wavefunction $G^{(2)}_{s,i}(\tau)$ expresses the time correlation between the signal and idler photons. By analyzing the full width at half maximum (FWHM) of $G^{(2)}_{s,i}(\tau)$, the correlation time $T_c$ and the bandwidth of the photon pairs can be determined\cite{Chuu_theory}. Furthermore, the normalized cross-correlation function $g^{(2)}_{s,i}=p_{s,i}/(p_sp_i)$ which quantifies the nonclassical feature of the photon pairs can be determined by normalizing $G_{s,i}^{(2)}(\tau)$\cite{sup}. The parameter $p_{s,i}$ is the probability of the coincidental detection of signal and idler photons and $p_{s}$($p_{i}$) is the probability of the detection of signal (idler) photons. 

\begin{figure}
\includegraphics[width=7.5 cm,viewport=30 75 830 530,clip]{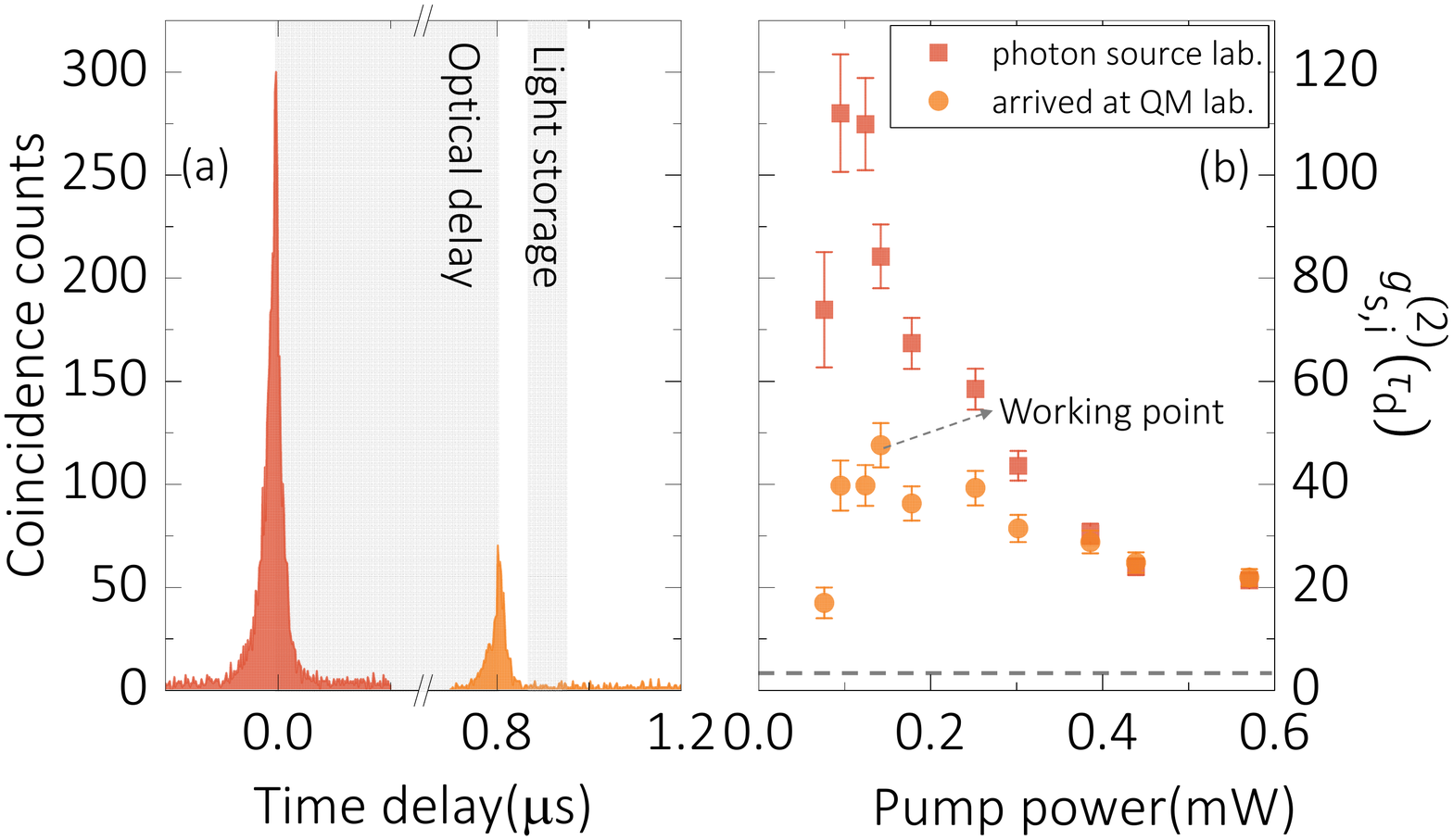}
\caption{(a) Biphoton wavefunction right after the photon source (red) and in the quantum memory laboratory (orange). The pump power is 0.14 mW. The $\sim$800 ns delay time in these two cases is due to the propagation delay in a 150 m fiber, an etalon filter and a $\sim$ 6 m free space path. (b) The normalized cross-correlation function versus the pump power right after the photon source (red squares) and in the quantum memory laboratory (yellow circles). The dashed line represent the classical limit of 2.}
\label{test}
\end{figure}

We measure $G_{s,i}^{(2)}(\tau)$ with idler photons just out of the source and with signal photons passing through a 150m long fiber to the quantum memory setup without loading the atoms. The travelling of the signal photons through the long fiber induces a delay time of $\sim 800$ns, which allows for a grace time window to compensate for the finite switching time of the function generator used to control the intensity of the coupling field through an acousto-optic modulator (AOM). The long fiber and other optical elements also induce an attenuation of the signal photons. The overall transmission efficiency of the signal photons in the quantum memory laboratory, without loading atoms, is around 25\%\cite{sup}. In order to operate the photon source at optimum conditions, the $g^{(2)}_{s,i}$ is measured for various pump powers, as shown in Fig.\ref{test}(b). The maximum $g^{(2)}_{s,i}$ appears at a pump power of 0.14 mW with values of $\approx84$ and $\approx47$ right after the photon source and in the quantum memory laboratory, respectively. The values are all above the nonclassical bound of $g^{(2)}_{s,i}=2$\cite{Kuzmich2003}. The corresponding detection rate at those two places are 1.10 and 0.24 kHz, respectively.

\begin{figure}
\includegraphics[width=8 cm,viewport=20 5 760 550,clip]{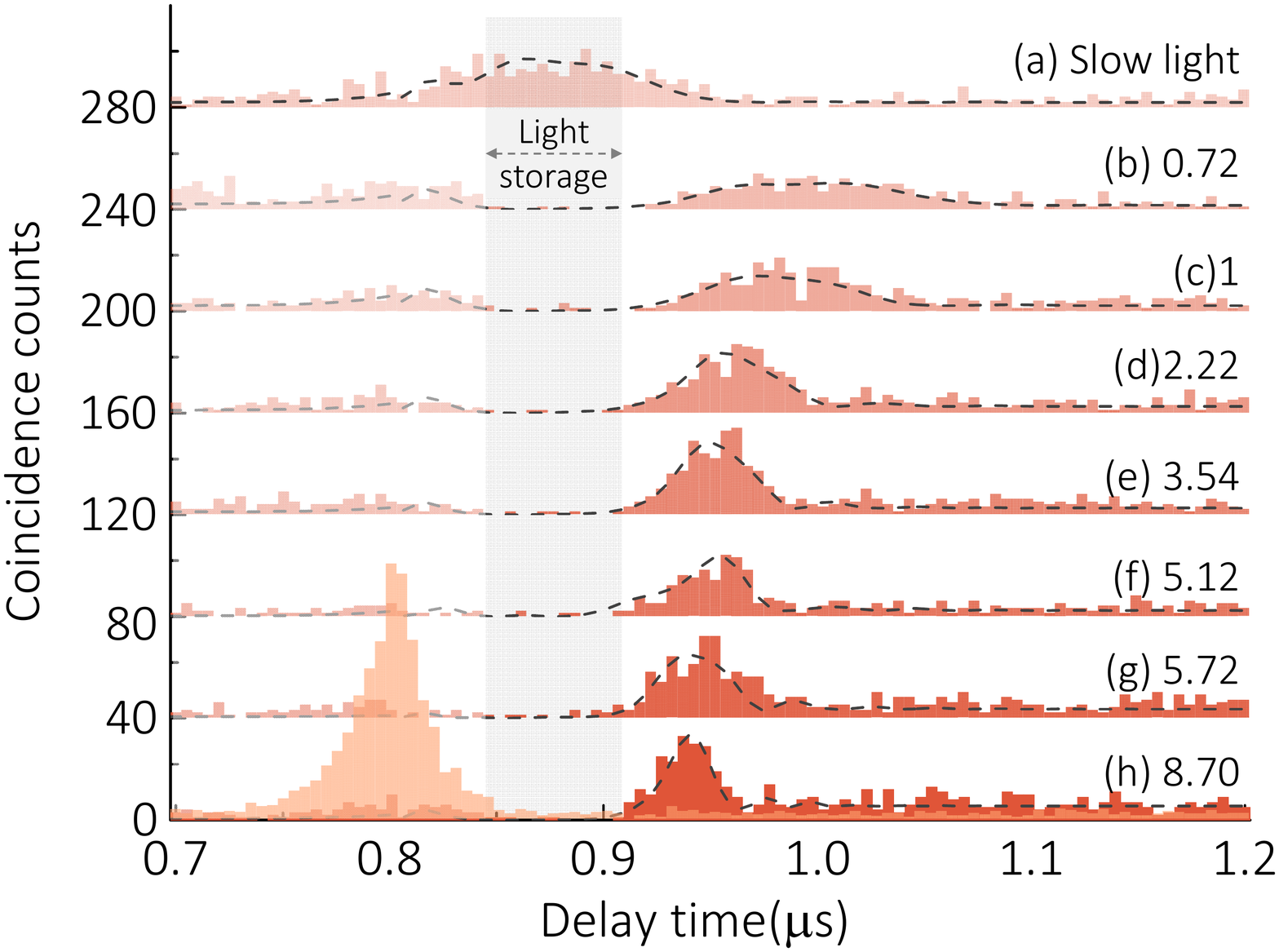}
\caption{Biphoton wavefunction for various coupling powers in the retrieval process. In case (a), the coupling beam is continuous on with a power of $P_w=$16 mW. This is the slow light case without storage. In all other cases, the signal photons are stored for 100 ns and retrieved with a different reading power for the coupling field. The power ratio between the reading to writing coupling field $\xi_{r/w}$ are also shown in the figure. The black dashed lines indicate the results of the theoretical simulation.}
\label{storage}
\end{figure} 

After characterizing the photon pairs, we then facilitate the storage of heralded single photons by loading of the cold atoms. Fig.\ref{setup}(c) depicts the timing sequence used in the experiment. Our cold atomic ensembles provide the EIT-memory with a repetition rate of 8 Hz. In each cycle, AOM$_1$ is on for 20 $\mu$s to release the signal photons 83 $\mu$s after turning off the MOT. In order to make sure that almost all of the biphoton wavefunction can be compressed into the EIT-memory, we first check whether the group delay time of the classical signal pulse is long enough\cite{YF2018}. The pump power of the photon source is raised up such that the signal output is in optical parametric oscillator (OPO) lasing mode and becomes a classical pulse of square waveform. This allows us to conveniently adjust the locking frequency of the signal photons to EIT resonance. We choose a writing power $P_w$ of 16 mW for the coupling beam such that the ratio of group delay time ($T_d$) of the signal pulse to the biphoton coherence time ($T_c\sim25$ns) is $\sim3$\cite{YF2018}. Once this is done, the photon source is set back to the photon pair mode by reducing the pump power. At each triggering of the idler photons, the coupling beam is turned off to store the signal photons. After a storage time of 100 ns, the coupling beam is turned on again to read out the signal photons. One representative example is shown in Fig.\ref{storage}(c). In this case, the reading power ($P_r$) of the coupling beam is equal to the writing power ($P_w=P_r$). The storage efficiency is $\sim$36$\%$, lower than the slow light efficiency of 52$\%$. The efficiency reduction is mainly due to the leakage of the front and real tails of the pulse. Due to the limited optical depth and thus the limited delay-bandwidth product\cite{DBP2007,YF2018}, the probe waveform is unavoidable to broaden significantly if one keeps a large enough group delay in order to compress the major part of the pulse into the medium\cite{YF2018}. 

There is a noticeable stretching of the biphoton wavefunction in the time domain, for both the slowed and the retrieved signal waveform with $P_r=P_w$, due to finite EIT-bandwidth. Effectively, the EIT medium narrows the spectral bandwidth of the heralded single photons and acts as a narrowband frequency filter. We estimate the bandwidth of the slow light to be 1.8 MHz by evaluating the reciprocal of the FWHM of $G^{(2)}_{s,i}(\tau)$. This bandwidth is much narrower than that of the input photons (6.2 MHz). In the case of the retrieved signal pulse with $P_r/P_w=1$, the determined bandwidth is 2.3 MHz. The slight bigger bandwidth compared to the slow light case is due to the leakage of the pulse tails.

\begin{figure}
\includegraphics[width=7.5 cm,viewport=20 55 800 510,clip]{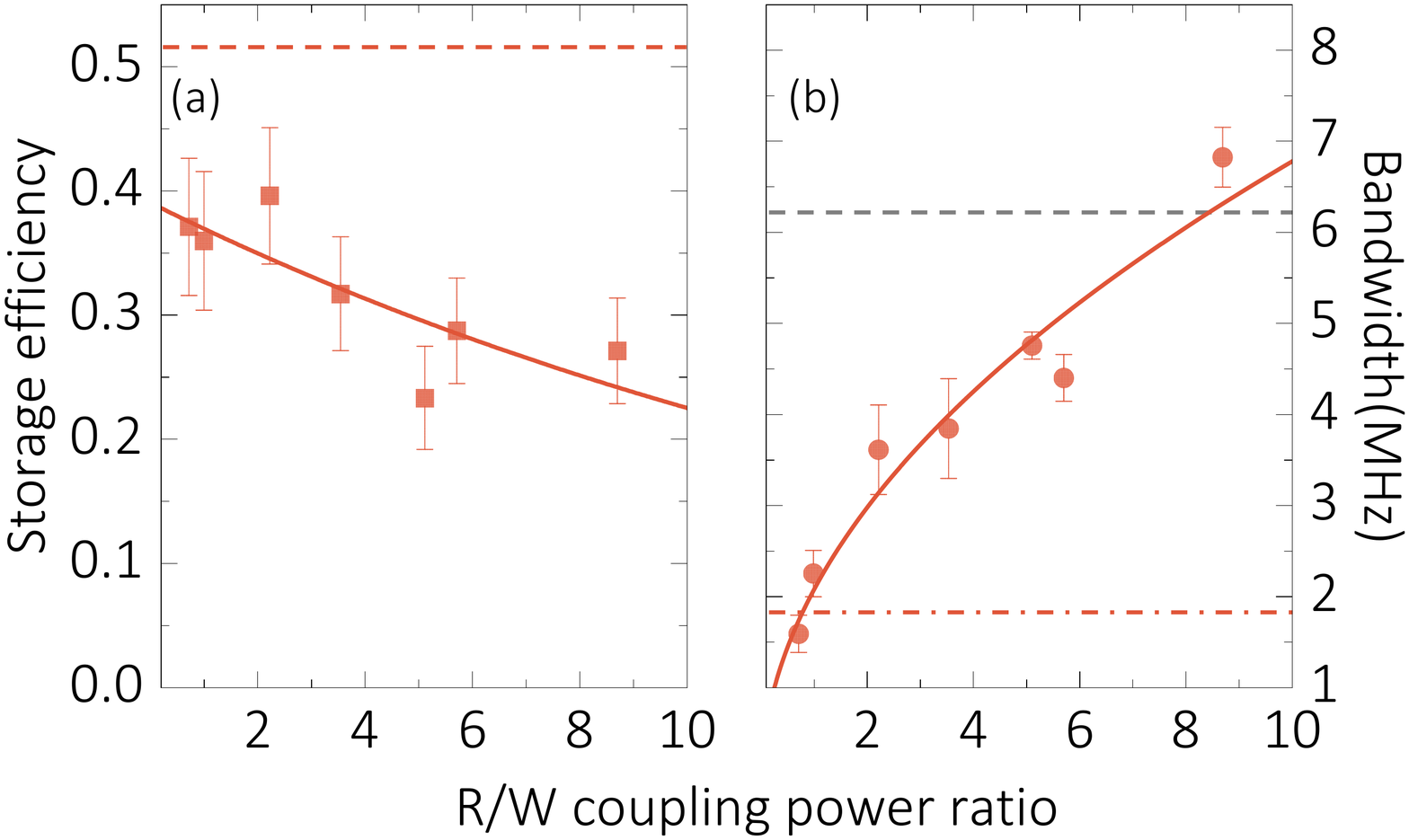}
\caption{(a) The storage efficiency versus the read to write (R/W) coupling power ratio $\xi_{r/w}$. The solid line is an exponential fitting of the data. The dashed line represents the efficiency of the slow light case. (b) The bandwidth of the biphoton wavefunction versus $\xi_{r/w}$. As $\xi_{r/w}$ increases, the bandwidth of the biphoton wavefunction increases. This demonstrates the manipulation of the heralded single photons. The solid line is a fitting curve with a relation of 2.12$\sqrt{\xi_{r/w}}$. The red dash dotted (gray dashed) line represents the bandwidth of the biphoton wavefunction for the slowed (input) photons.}
\label{eff_bw}
\end{figure}    

In\cite{sup}, we quantitatively discuss the evolution of the biphoton wavefunction $G^{(2)}_{s,i}(\tau)$ after passing through an EIT memory based on the Heisenberg-Langevin equation. The physical meaning of the result is clear if the problem is analyzed in the frequency domain. The conclusion is that $G^{(2)}_{s,i}(\tau)$ is just its initial frequency component times the EIT medium response function and transformed back to the time domain. There are two additional terms contributing to the background of $G^{(2)}_{s,i}(\tau)$ in the theory. One is due to the accidental measurement of the photon pairs. The other is due to the spontaneous decay, which is negligible at the strong coupling limit. In addition, technical factors such as coupling leakage, stray light leakage and the dark count of the single photon counter may also contribute to the background of $G^{(2)}_{s,i}(\tau)$. Equivalently, $G^{(2)}_{s,i}(\tau)$ can be calculated by solving the Maxwell-Bloch equations with its initial temporal waveform as the input. The black dashed lines in Fig.\ref{storage} indicate the numerical calculations to model the data obtained by this way. From numerical fit to the slow light data in Fig.\ref{storage}(a), we estimate an overall ground-state decoherence rate $\gamma_d$ of $(0.065\pm0.01)\Gamma$. Multiple factors may contribute to the ground-state decoherence, including the residual Doppler broadening\cite{PanJW2009,Yu2011}, inhomogeneous stray magnetic field, finite mutual laser coherence between the signal and coupling fields, and coupling-intensity-dependent decoherence due to off-resonant excitation to the nearby cycling transition\cite{YF2018}. 

EIT-memories have been used to manipulate the temporal width and other properties of classical signal pulses by varying the coupling parameters during the retrieval process\cite{HakutaPRA,Yu2005,Yu2006}. These ideas can be extend to the quantum regime to manipulate the biphoton wavefunction $G^{(2)}_{s,i}(\tau)$. In this work, we vary the coupling reading powers $P_r$ to manipulate the retrieved biphoton waveform. The reading to writing coupling power ratio is denoted as $\xi_{r/w}=P_r/P_w$. In an ideal EIT memory, the retrieval efficiency is only determined by the writing process, independent of the reading power of the coupling\cite{OptimalStorage}. By choosing $\xi_{r/w}>1$, the retrieved $G_{s,i}^{(2)}(\tau)$ can be compressed and the peak value of $G^{(2)}_{s,i}(\tau)$ can be enhanced.

\begin{figure}
\includegraphics[width=6.5 cm,viewport=40 80 640 465,clip]{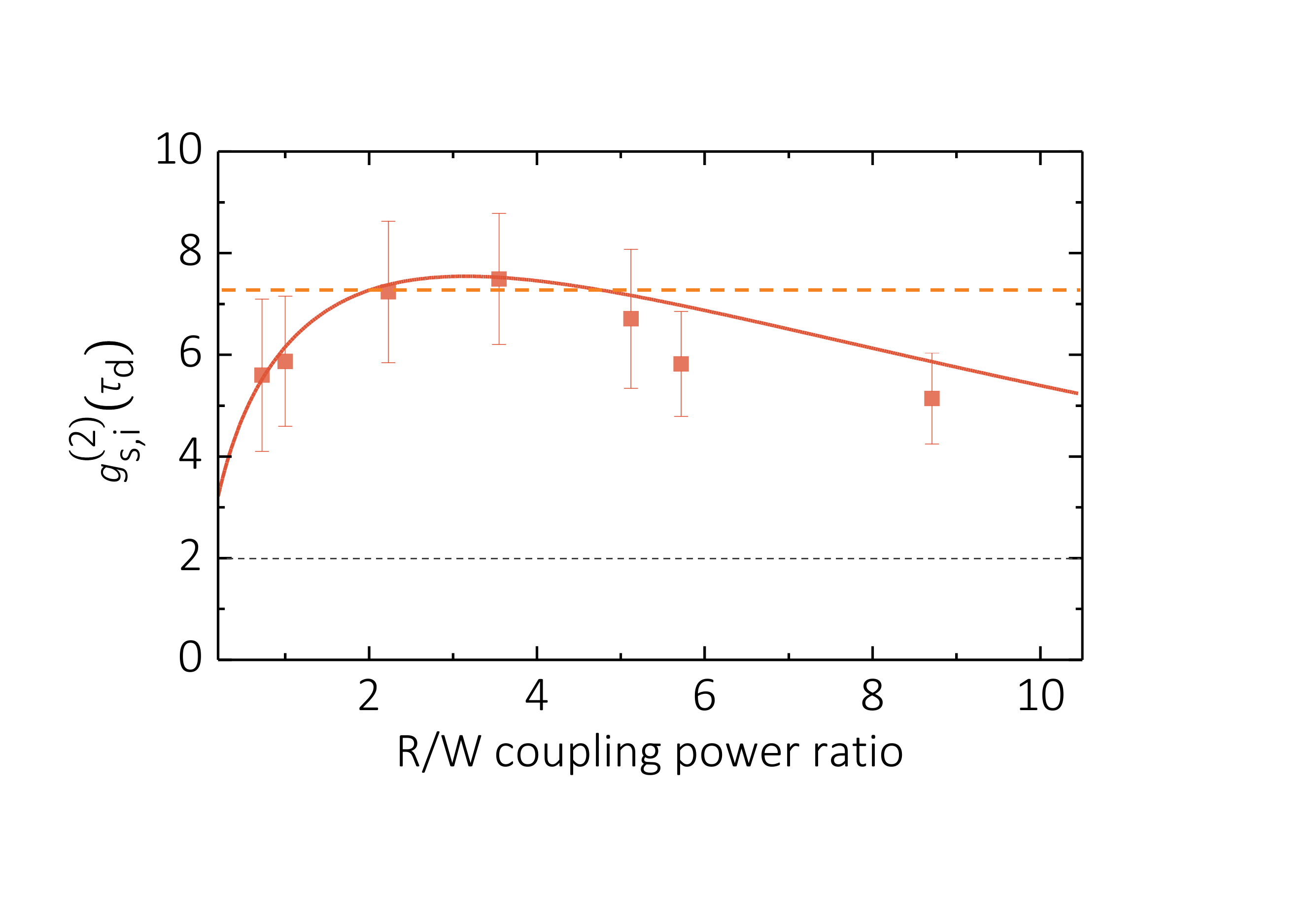}
\caption
{Manipulation of non-classical correlation of the retrieved photons. The red squares represent the $g^{(2)}_{s,i}(\tau_d)$ of the retrieval photons with different R/W power ratios. The red solid line shows the theoretical fit based on Eqs.\ref{g2_mod} with the parameters $\gamma_s\sim0.055$, $\alpha=0.43$ and $N_b=2.8$. The red dashed line is the cross-correlation of the slow light case. The gray dashed line depicts the classical limit of $g_{s,i}^{(2)}=2$.
}
\label{g2}
\end{figure} 

We utilize a series of $P_r$ to retrieve the heralded single photons, as shown in Fig.\ref{storage}. By increasing $P_r$ (Fig.\ref{storage} (c)$\sim$(g)), the biphoton wavefunctions are compressed due to the increase of group velocities during the reading phase. An opposite case with $\xi_{r/w}<1$ is shown in Fig.\ref{storage}(b), in which $G_{s,i}^{(2)}$ broadens. In Fig.\ref{eff_bw}(b), we demonstrate the bandwidth control of the heralded single photons based on the data of Fig.\ref{storage}. The corresponding storage efficiencies are shown in Fig.\ref{eff_bw}(a), indicating an exponential decay relation versus $\xi_{r/w}$ but not a constant relation as expected in an ideal EIT system\cite{OptimalStorage}. This is due to the off-resonant excitation of the coupling field to the nearby excited state (e.g. 6$P_{3/2}, F=5$ state), which induces a multiphoton decay channel and thus a coupling-intensity-dependent ground-state decoherence rate\cite{YF2018}. If the EIT memory is operated in cesium $D_1$ line, this problem will be significantly reduced\cite{YF2018}. The storage efficiency versus $\xi_{r/w}$ of Fig.\ref{eff_bw}(a) can be modeled as $e^{-\gamma_s \xi_{r/w}}$\cite{YF2018}, where $\gamma_s$ is 0.055 as determined by exponential fitting of the data. Theoretically, $\gamma_s$ is related to the optical depth, Rabi frequency of the coupling and the coupling detuning with respect to the $6P_{3/2}, F=5$ state\cite{YF2018}. By plugging in these parameters, we check that the calculated value is consistent with the fitted parameter to within 10\%. The observed bandwidth of the retrieved photons is proportional to $\sqrt{\xi_{r/w}}$, instead of $\xi_{r/w}$\cite{EITbandwidth}. This trend is understandable and is explained below. There is a gradual shift in the transparent bandwidth of an EIT system versus the coupling intensity from a linear to a square root relation as it increases\cite{sup}. The bandwidth of our photon pairs is relatively wide such that we operate the coupling in the high intensity regime, which explains the $\sqrt{\xi_{r/w}}$ trend for the bandwidth of the retrieved photons. 

We next address the cross-correlation measurement of the photon pairs, $g^{(2)}_{s,i}$. For the slowed and retrieved cases with $\xi_{r/w}$ between 0.72 and 8.7, the peak $g^{(2)}_{s,i}$ are all larger than the nonclassical threshold value of 2\cite{Kuzmich2003}, as shown in Fig.\ref{g2}. Therefore, the nonclassical nature of the photon pairs is preserved in the EIT-memories. Nevertheless, these values of $g^{(2)}_{s,i}$ are significantly lower than that of the input photons, which is 47. The effects that cause the efficiency reduction also cause a reduction of the $g^{(2)}_{s,i}$. One important technical factor that leads to a reduction of $g^{(2)}_{s,i}$ is the leakage of the coupling beam into the single photon counter. To minimize this leakage, we introduce an angle of $4^0$ between the coupling and probe beam and use four irises along the signal beam path, as well as one etalon filter before the detector\cite{sup}. The overall extinction ratio of the coupling beam is 129 dB. Increasing the storage efficiency of the memory, minimizing the technical losses of all optical components, and increasing the extinction ratio of the filtering of coupling beam will enhance the nonclassical feature.

As the coupling power during retrieval increases, the peak $g_{s,i}^{(2)}(\tau)$ first increases, until reaching a maximum and then decreases. The maximum $g_{s,i}^{(2)}(\tau) of \sim7.5$ appears at $\xi_{r/w}\sim3.5$, which is greater than 5.8 for the case of $\xi_{r/w}=1$. This demonstrates that the nonclassical feature of the photon pairs can be enhanced by manipulating the coupling power during retrieval. The enhancement ratio depends on multiple factors. As discussed previously, it should scale as $\sqrt{\xi_{r/w}}$ at the high coupling intensity regime. Due to the coupling-intensity-dependent decoherence rate, there is an additional factor $e^{-\gamma_s \xi_{r/w}}$. The accidental photons from the photon source and the dark count of the detector contribute to the background of the coincidence count (denoted as $N_b$ for a given integration time) and thus the degradation of $g_{s,i}^{(2)}(\tau)$. The leakage of photons from the coupling beam into the detector also contributes to the background coincidence count. This leakage rate is proportional to the coupling power and we denote it to be a proportional constant $\alpha$ times $\xi_{r/w}$. Based on these discussions, we write down the relation for the peak $g_{s,i}^{(2)}(\tau)$ versus $\xi_{r/w}$ as follows,
\begin{equation}
g^{(2)}_{s,i}(\tau_d)=\frac{N_{s,i}\sqrt{\xi_{r/w}}e^{-\gamma_s \xi_{r/w}}}{\alpha\xi_{r/w}+N_{b}},
\label{g2_mod}
\end{equation}
where $N_{s,i}$ ($N_{b}$) denotes the coincidence (background) count within a given integration time and $\tau_d$ denotes the group delay time at which $g^{(2)}_{s,i}$ reaches its peak value. As shown in Fig.\ref{g2}, the fit curve based on this relation is in good agreement with the experimental data.

In summary, we demonstrate the quantum storage and manipulation of heralded single photons generated from a single-mode cavity-SPDC setup. The bandwidth and the nonclassical correlation of the photon pairs can be manipulated by varying the coupling intensity during the retrieval process. Despite some imperfect effects, the nonclassical correlation of the retrieved photons can be enhanced and even surpass that of the slow light case. Significant improvement is possible for operation of the photon source and EIT memory in the $D_1$ transition\cite{YF2018}. Our experimental interface between a single-mode cavity-SPDC photon source and atomic quantum memories lays the foundation for the realization of important quantum information protocol, such as quantum repeaters and enhancing the multiphoton generation rate.  

This work was done as part of a collaborative project under the Science Vanguard Program and Academic Summit Program of Ministry of Science and Technology (MOST) of Taiwan, with Ite A. Yu as the project leader and Y.-C. Chen, Y.-F. Chen and others as the subproject leaders. We thank I. A. Y. and Y. F. C. for their fruitful discussions. We also thank Yen-Yin Li and Sheng-Lung Huang for technical assistance with fiber fusion. This work was supported by MOST of Taiwan under Grant No. 106-2119-M-001-002, 107-2112-M-001-003, and 107-2745-M-007-001. We also thank the supports from NCTS ECP1 and the Center for Quantum Technology from the Featured Areas Research Center Program of Ministry of Education of Taiwan. 

P.-J. T. and Y.-F. H. contributed equally to this work.

%



\pagebreak

\onecolumngrid
\begin{center}
  \textbf{\large Quantum storage and manipulation of heralded single photons in atomic quantum memories: Supplementary Material}\\[.2cm]
  Pin-Ju Tsai$^{1,2}$, Ya-Fen Hsiao$^{2,3}$, and Ying-Cheng Chen$^{2}$\\[.1cm]
  {\itshape ${}^1$Department of Physics, National Taiwan University, Taipei 10617, Taiwan\\
  ${}^2$Institute of Atomic and Molecular Sciences, Academia Sinica, Taipei 10617, Taiwan\\
  ${}^3$Molecular Science Technology Program, Taiwan International Graduate Program, Academia Sinica and National Central University, Taiwan\\}
  ${}^*$chenyc@pub.iams.sinica.edu.tw\\
\end{center}


\maketitle
\section{Experimental setup}
\subsection{Preparation of EIT-based quantum memory}
In the quantum memory laboratory, two techniques are used to increase the optical depth of the cold atomic media. First, we utilize the temporal dark MOT to increase the atom density, in which the power of the MOT repumping beam is reduced from 12 mW to 1 mW for 5 ms to shelve most of the population into the dark hyperfine ground state, $\left|6S_{1/2},F=3\right\rangle$. Second, a Zeeman pumping beam driving the $\left|6S_{1/2},F=3\right\rangle\rightarrow\left|6P_{3/2},F=2\right\rangle$ $\sigma^{+}$-transition is used to pump most of the population into the rightmost Zeeman state in which the $\sigma^{+}$ signal transition has the largest transition dipole moment. The Zeeman pumping beam is turned on for 63$\mu$s with an intensity of 13.8 mW/cm$^2$. To measure the optical depth, we use a coupling beam and a weak signal beam to drive the $\left|6S_{1/2},F=4\right\rangle\rightarrow\left|6P_{3/2},F=4\right\rangle$ and $\left|6S_{1/2},F=3\right\rangle\rightarrow\left|6P_{3/2},F=4\right\rangle$ transitions, respectively. By scanning the frequency of the signal beam, we can measure the transmission spectrum and determine the optical depth of the EIT medium by spectral fitting. Typical optical depths in the experiment are about $55\pm10$. 

\subsection{Operation of the photon-pair source}
The pump power of our cavity-enhanced SPDC photon-pair source alternates between a high power ($\sim$40 mW) and a low power phase (0.1$\sim$0.6 mW) at a repetition rate of 6.7 kHz,  controlled by an AOM. In the high power phase, the output of the source is in lasing mode (also called the optical parametric oscillator, OPO). In the low power phase, the output is in photon pair mode. We use the OPO output light to lock the cavity into the double-resonant condition. Due to the unbalanced reflectivity of the two cavity end mirrors, about 60\% (40\%) of the OPO power is reflected back (is transmitted through) the cavity (Fig.1(a) of the main text). After splitting the signal (852nm) and idler (780 nm) components in the transmitted OPO light with a polarizing beam splitter (PBS), the idler light is detected by a photodetector which acts as a power monitor for the OPO light. After the PBS, the transmitted signal light is further split into two parts by a 50/50 beam splitter. One part of the light is detected by a photodetector and the output is used to obtain a demodulated error signal to lock the cavity to resonance through the piezo attached to the cavity output coupler. Another part of the signal light interferes with a reference laser which is locked to the cesium D$_2$-line $|6S_{1/2},F=3\rangle\rightarrow|6P_{3/2},F=4\rangle$. With this beatnote, we lock the frequency of the signal light at the desired atomic transition using the simple offset frequency locking scheme \cite{locking}. During the photon pair output phase, the photon pairs emitted in the backward direction from the cavity are picked up by a dichroic mirror, diffracted by AOM$_1$, and split apart by a PBS. The idler photons are detected by a single photon counter SPCM$_i$ and the signal photons are coupled into a long fiber and directed to the quantum memory laboratory.

\subsection{Filtering setup and collection efficiency}
It is crucial to filter out the coupling beam and other unwanted stray light during the single-photon storage experiment. Multiple techniques are used to minimize the background photons including the introduction of a $4^0$ angle between the coupling and signal beams and blocking the coupling beam by a beam trap. After passing through the MOT cell, the signal beam propagates along a long beam path ($\sim$6 m) with four irises to further block the coupling light. The signal beam then passes through a bandpass filter (Sermock LL01-852-12.5) and an etalon filter (Quantaser FPE001). After coupling into a single-mode fiber, the signal beam is then detected by the single-photon counter. The overall extinction ratio for the coupling beam is 129.1 dB. The overall collection efficiency of the signal light is 25.0\%, which is defined as the power ratio between that on the single-photon counter and the output after passing through the 150m fiber. Details of the transmission efficiency for the components along the beam path include 76\% for a polarizer, 89.4\% for some lenses and the MOT cell, 51.0\% for the etalon, 90\% for the bandpass filter and 80\% for the coupling of the single-mode fiber.

\subsection{Measurement of the coincidence count and cross-correlation function}
A digital oscilloscope (Rohde \& Schwarz RTO-1014) was used for the coincidence measurement and collection of data statistics. As mentioned in the main text, a 150m long fiber was used to induce a $\sim$ 800 ns propagation delay for the signal photons. We chose a total detection time window of 500 ns, starting at 700 ns, after being triggered by the idler photons. The coincidence histograms are shown in Fig. 3 of the main text. This is a 100-bin histogram with a 5 ns time resolution. To measure $g_{s,i}^{(2)}(\tau_d)$ which is evaluated by $p_{s,i}/(p_{s}p_{i})$, we observe the number of coincidences $N_{s,i}(\tau_d)$ at the peak of the biphoton wavefunction (with a delay time $\tau_d$ ). The uncorrelated coincidence count $N_{s}$ is determined by estimating the coincidence count far away from the biphoton wavefunction. Since the conditional measurement is already normalized by $p_{i}$, the cross-correlation function can be evaluated by  
\begin{equation}
g_{s,i}^{(2)}(\tau_d)=\frac{N_{s,i}(\tau_d)}{\langle N_{s}\rangle},
\label{g2mea}
\end{equation}
where $\langle N_{s}\rangle$ denotes the average of the uncorrelated coincidence counts. The coincidence histogram in Fig. 3 of the main text is a statistics of 30000 idler trigger events. With the 8 Hz repetition rate of the experiment, the actual time for data collection is around 1.5 hours. 

\section{Theoretical background}

\subsection{Theoretical background of EIT-quantum memory}
In this section, we consider the full quantum theory of a quantized signal field propagating through an EIT-quantum memory (QM). As shown in Fig.1(b) of the main text, the EIT-QM is constituted of a three-level atomic system that contains two ground states of $|F=3(4)\rangle=|g(s)\rangle$ and one excited state $|F'=4\rangle=|e\rangle$. The atomic population is prepared at $|g\rangle$ initially. A strong, classical light couples the $|s\rangle\leftrightarrow|e\rangle$ transition with a Rabi frequency of $\Omega_c$. A quantized signal field $\hat{a}_s(z,t)$ is on resonance with the $|g\rangle\leftrightarrow|e\rangle$ transition. The equation of motion for the atomic system is given by the Heisenberg-Langevin equations,
\begin{equation}
\frac{\partial}{\partial t}\hat{\sigma}_{ge}(z,t)=\frac{i}{2}\Omega_c\hat{\sigma}_{gs}(z,t)+\frac{i}{2}g_s\hat{a}_s(z,t)+\left(i\delta_{ge}-\frac{\Gamma}{2}\right)\hat{\sigma}_{ge}(z,t)+\hat{F}_{ge},
\label{si31eq}
\end{equation}
\begin{equation}
\frac{\partial}{\partial t}\hat{\sigma}_{gs}(z,t)=\frac{i}{2}\Omega_c^*\hat{\sigma}_{gs}(z,t)+\left(i\delta_{gs}-\frac{\gamma_{gs}}{2}\right)\hat{\sigma}_{gs}+\hat{F}_{gs},
\label{si21eq}
\end{equation}
where $\delta_{ge}$ and $\delta_{gs}$ denote the one- and two-photon detuning of the signal field, $\hat{\sigma}_{ij}$ is the slowly varying atomic coherence operator, $\hat{F}_{ij}$ is the Langevin noise operator, $g_s$ is the coupling constant for atom-field interaction, $\Gamma$ is the spontaneous decay rate from $|e\rangle$ to ground states and $\gamma_{gs}$ is the decoherence rate between $|g\rangle$ and $|s\rangle$. The equation of motion for the signal field propagating in the atomic medium is described by the Maxwell-Schrodinger equation of 
\begin{equation}
\left(\frac{1}{c}\frac{\partial}{\partial t}+\frac{\partial}{\partial z}\right)\hat{a}_s(z,t)=\frac{ig_sN}{c}\hat{\sigma}_{ge},
\label{MSE}
\end{equation}
where $N$ is the atom number and $c$ is the speed of light in a vacuum. 
To solve the problem, we first Fourier transform Eq.\ref{si31eq}-\ref{MSE} to the frequency domain. By solving the coupled equations of Eqs.\ref{si31eq}-\ref{si21eq}, the atomic coherence can be obtained as follows,

\begin{equation}
\hat{\sigma}_{ge}(\omega)=\frac{ig_s}{2}\frac{d_{gs}}{\mathbf{D}(\omega)}\hat{a}_s+\hat{f}(\omega),
\label{s13}
\end{equation}
where $\mathbf{D}(\omega)=d_{ge}d_{gs}+|\Omega_c/2|^2$, $d_{ij}=\gamma_{ij}/2-i(\omega+\delta_{ij})$. And $\hat{f}(\omega)=(i\Omega_c\hat{F}_{gs}+d_{gs}\hat{F}_{ge})/\mathbf{D}(\omega)$.
By substituting Eq.\ref{s13} into Eq.\ref{MSE}, the field equation become
\begin{equation}
\begin{aligned}
&\frac{\partial}{\partial z}\hat{a}_s-\frac{i\omega}{c}\hat{a}_s=-\frac{g_s^2N}{2c}\frac{d_{gs}}{\mathbf{D}(\omega)}\hat{a}_s+\frac{ig_sN}{c}\hat{f}(\omega),
\end{aligned}
\end{equation}
and
\begin{equation}
\begin{aligned}
\hat{a}_s(z,\omega)=&\hat{a}_s(0,\omega)e^{-\Lambda(\omega)z}+\delta \hat{a}_s(z,\omega),\\
\delta \hat{a}_s(z,\omega)=&\int_0^z\frac{ig_sN}{c}\hat{f}(\omega)e^{\Lambda(\omega)(z'-z)}dz,
\label{a_eit}
\end{aligned}
\end{equation}
where $\Lambda(\omega)=-\frac{i\omega}{c}+\frac{g_s^2Nd_{gs}}{2c\mathbf{D}(\omega)}$. Eq.\ref{a_eit} shows the evolution of the signal field after interacting with the EIT medium in the frequency domain, which contains two terms. The first term is just the input signal field multiplied by the EIT-medium frequency response. This term represents the transmissive and dispersive proprieties of the signal field during its propagation in an EIT medium. The second term $\delta\hat{a}_s(z,\omega)$ in the field solution describes the field fluctuations from the atomic system. The field fluctuations can be neglected under the strong coupling condition since the  $\delta\hat{a}_s(z,\omega)$ is contributed by the atomic population in the excited state\cite{PhysRevA.46.5856,PhysRevA.88.013823}. 


Eq.\ref{a_eit} provides the solution depicting the interaction between the field and EIT-QM. By selecting an initial quantum field of $\hat{a}_s(0,\omega)$, we can simulate the output fields from the EIT QM. In our case, the signal field is the single photon generated by the cavity-enhanced SPDC. In the next section, we write down the field solution for the cavity-SPDC to evaluate the interaction between the SPDC-field and EIT QM.

\subsection{Biphoton wavefunction in an EIT medium}
For a cavity-enhanced SPDC under single-mode operation, the signal and idler output fields are given by\cite{PhysRevA.83.061803}

\begin{equation}
\begin{aligned}
\hat{a}_{s}^{out}(\omega)&=A_s(\omega)\hat{a}_{s}^{in}(\omega)+B_s(\omega)\hat{a}_{i}^{in\dagger}(-\omega_i),\\
\hat{a}_{i}^{out\dagger}(-\omega_i)&=B_i(\omega)\hat{a}_{s}^{in}(\omega)+A_i(\omega)\hat{a}_{i}^{in\dagger}(-\omega_i),
\label{field}
\end{aligned}
\end{equation}
where $\hat{a}(\omega)=\frac{1}{2\pi}\int\hat{a}(t)exp(i\omega t)dt$ is the Fourier transform of the field operator from the time to the frequency domain, and $\omega_i=\omega_{pump}-\omega$. The coefficients in Eq.\ref{field} are

\begin{equation}
\begin{aligned}
A_{s,i}(\omega)&=\frac{\gamma_{s,i}-\Gamma_{s,i}/2+i(\omega_{,i}-\Omega_{q,r})}{\Gamma_{s,i}/2-i(\omega_{,i}-\Omega_{s,r})},\\
B_{s,i}(\omega)&=\frac{\mp i\kappa\sqrt{\gamma_s\gamma_i}}{[\Gamma_s/2-i(\omega-\Omega_q)][\Gamma_i/2+i(\omega_i-\Omega_r)]},\\
\end{aligned}
\end{equation}
where $\Omega_{q,r}$ is the frequency of the cavity mode of the signal and idler field\cite{PhysRevA.83.061803}. For the double resonance condition of the cavity, we have a relation of $\Omega_{q}+\Omega_{r}=\omega_{pump}$ and thus $\omega_i-\Omega_r=\Omega_q-\omega$.

In the experiment the signal photons are sent into the EIT medium. The field operator of the input signal field $\hat{a}_s(0,\omega)$ in Eq.\ref{a_eit} can be replaced by the signal field operator out of the photon source, i.e. $\hat{a}_s^{out}(\omega)$. By replacing $\hat{a}_s(0,\omega)$ with $\hat{a}_s^{out}(\omega)$ and combine Eq.\ref{field}, we obtain the signal field operator in the EIT medium as follows,

\begin{equation}
\begin{aligned}
\hat{a}_s^{EIT}(z,\omega)=\alpha_s(z,\omega)\hat{a}_{s}^{in}(\omega)+\beta_s(z,\omega)\hat{a}_{i}^{in\dagger}(-\omega_i),
\label{a_eit_1}
\end{aligned}
\end{equation}
where $\alpha_s(z,\omega)=A_s(\omega)e^{-\Lambda(\omega)z}$. $\beta_s(z,\omega)=B_s(\omega)e^{-\Lambda(\omega)z}$. In the experiment, we measure the two-photon correlation function which is 
\begin{equation}
\begin{aligned}
G^{(2)}_{s,i}(t_s,t_i)=\langle \hat{a}_i^{out\dagger}(t_i)\hat{a}_s^{out\dagger}(t_s)\hat{a}^{out}_p(t_s)\hat{a}^{out}_i(t_i) \rangle,
\end{aligned}
\end{equation}
 where $t_{s,i}$ is the time when the signal (idler) photon hits the detector. Using Eqs.\ref{field} and \ref{a_eit_1}, the two-photon correlation function for signal photons passing through the EIT medium is thus, 
\begin{equation}
\begin{aligned}
G^{(2)}_{s,i}(\tau)=&\left|\frac{1}{2\pi}\int_{-\infty}^{\infty}d\omega \alpha_s^{\dagger}(z,\omega)B_i(\omega)e^{i\omega\tau}\right|^2+\langle n\rangle,
\label{G22_EIT}
\end{aligned}
\end{equation}
where 
\begin{equation}
\begin{aligned}
&\langle n\rangle=\frac{1}{(2\pi)^2}\int_{-\infty}^{\infty}d\omega|B_i(\omega)|^2\times\\
&\left[ \int_{-\infty}^{\infty}d\omega|\beta_s(z,\omega)|^2+\int_{-\infty}^{\infty}\int_{-\infty}^{\infty}d\omega d\omega'\langle\delta\hat{a}^{\dagger}_s(z,\omega)\delta\hat{a}_s(z,\omega') \rangle e^{i(\omega'-\omega)t_s}\right ].
\label{noise}
\end{aligned}
\end{equation}
The second term in Eq.\ref{G22_EIT} is a time-independent background quantity which contains two terms, as shown in Eq.\ref{noise}. The first term is due to the accidental coincidence measurements with nearby photons, which can be suppressed by choosing a lower pump field to avoid too many photon pairs being generated. The second term is due to the spontaneous decay which is negligible under the strong coupling condition. Technically, the leakage of the coupling photons, the stray light photons, and the dark count of the detector may also contribute to the background count. The main property of the photon-atom interaction is determined by the first term of Eq.\ref{G22_EIT}. This term is nothing but the Fourier transform of the frequency component of the waveform $A_s^{\dagger}({\omega})B_s(\omega)$ multiplied by the EIT medium frequency response $e^{-\Lambda(\omega)z}$ back to the time domain. To simulate the biphoton wavefunction after passing through an EIT medium, it can be calculated in the frequency domain by Eq.\ref{G22_EIT} with an initial waveform of $A_s^{\dagger}({\omega})B_s(\omega)$, or it can be calculated using Eqs.\ref{si31eq}-\ref{MSE} with the input pulse of $A_s^{\dagger}({\omega})B_s(\omega)$ in the time domain. 

\subsection{Bandwidth of an EIT medium}
In Supplementary Note 3, we consider the bandwidth of an EIT medium versus the coupling intensity. Considering the semi-classical model of Eq.\ref{a_eit}, we obtain the transmission of the signal field after passing through an EIT medium by,
\begin{equation}
\begin{aligned}
T(\delta_{ge})&=\exp\left\{-2Re[\Lambda(0)]L\right\}\\
&=\exp\left[-\alpha\Gamma\frac{\gamma_{gs}|\Omega_c|^2+(4\delta_{gs}^2+\gamma_{gs}^2)\Gamma}{(|\Omega_c|^2+\Gamma\gamma_{gs}-4\delta_{ge}\delta_{gs})^2+(2\delta_{ge}\gamma_{gs}+2\delta_{gs}\Gamma)^2}\right],
\label{T_eit}
\end{aligned}
\end{equation}
where $\alpha=2NLg_s^2/\Gamma c$ is the optical depth. Eq.\ref{T_eit} is the EIT-spectrum as a function of one- and two-photon detuning. In a typical experiment, the coupling detuning is set to zero and thus $\delta_{gs}=\delta_{ge}$. Fitting the EIT spectrum to Eq.\ref{T_eit} allows us to determine the experimental parameters, with one representative example shown in Fig.1d. Based on Eqs.\ref{T_eit}, we can determine the full width at half maximum (FWHM) of the EIT transparent bandwidth, which is
 \begin{equation}
\begin{aligned}
\Delta\omega_{EIT}=\sqrt{(x+y)\left[1-\left(1-\frac{y^2}{(x+y)^2}\right)^{\frac{1}{2}}\right]}\Gamma,
\label{linewidth_EIT}
\end{aligned}
\end{equation}
where $x=\alpha/(2\ln2)-1/2$ and $y=|\Omega_c|^2/\Gamma^2$. There is a lower limit for the optical depth of $\ln2$ such that the factor $x$ in Eq.\ref{linewidth_EIT} is greater than zero. This is because the absorption depth needs to surpass 0.5 in order to possibly define the FWHM bandwidth of the transparent window. Based on Eq.\ref{linewidth_EIT}, we can roughly divide the $\Delta\omega_{EIT}$ into two regimes of $x\gg y$ and $x\ll y$, depending on the intensity of the coupling field. For the case of low coupling intensity ($x\gg y$), the EIT transparent bandwidth is 
 \begin{equation}
\begin{aligned}
\Delta\omega_{EIT}\approx\frac{y}{\sqrt{2x}}\Gamma
=&\sqrt{\frac{\ln2}{\alpha}}\frac{|\Omega_c|^2}{\Gamma},
\label{linewidth_1}
\end{aligned}
\end{equation}
which is linearly proportional to the coupling intensity and inversely proportional to $\sqrt{\alpha}$. For the case of high coupling intensity ($y\gg x$), the EIT transparent bandwidth is 
 \begin{equation}
\begin{aligned}
\Delta\omega_{EIT}\approx|\Omega_c|(1-\sqrt{\frac{x}{2y}})\approx|\Omega_c|.
\label{linewidth_2}
\end{aligned}
\end{equation}
Eq.\ref{linewidth_2} shows that the EIT transparent bandwidth is approximately proportional to the Rabi frequency of the coupling field in the high coupling intensity regime. This explains the $\sqrt{\xi_{r/w}}$ scaling law for $g_{s,i}^{(2)}(\tau_d)$ in Eq.1 of the main text. The dependence of the EIT bandwidth on the coupling intensity is a key point for bandwidth control in the biphoton wavefunction.  

\twocolumngrid

\end{document}